\documentclass[submission,copyright,creativecommons]{eptcs}
\providecommand{\event}{F-IDE 2015}
\usepackage{xspace,url,color,graphicx}

\newcommand{\ocaml}{\textsl{OCaml}\xspace}
\newcommand{\framac}{\textsl{Frama-C}\xspace}
\newcommand{\cil}{\textsl{CIL}\xspace}
\newcommand{\acsl}{\textsl{ACSL}\xspace}
\newcommand{\Value}{\textsl{Value}\xspace}
\newcommand{\WP}{\textsl{WP}\xspace}
\newcommand{\eacsl}{\textsl{E-ACSL}\xspace}
\newcommand{\rte}{\textsl{RTE}\xspace}
\newcommand{\C}{\textsl{C}\xspace}
\newcommand{\cpp}{\textsl{C++}\xspace}
\newcommand{\java}{\textsl{Java}\xspace}
\newcommand{\eclipse}{\textsl{Eclipse}\xspace}
\newcommand{\git}{\textsl{Git}\xspace}

\title{Software Architecture of\\ Code Analysis Frameworks
  Matters:\\ The Frama-C Example}

\def\titlerunning{Software Architecture of Code Analysis Frameworks Matters}
\def\authorrunning{J. Signoles}

\author{Julien Signoles\thanks{This work has been partly supported
    by the European FP7 project Stance 317753.}
\institute{CEA LIST, Software Security Lab, \\
  PC 174, 91191 Gif-sur-Yvette, France\\
  \email{julien.signoles@cea.fr}}}

\begin{document}

\maketitle


\begin{abstract}
Implementing large software, as software analyzers which aim to be used in
industrial settings, requires a well-engineered software architecture in order
to ease its daily development and its maintenance process during its
lifecycle. If the analyzer is not only a single tool, but an open extensible
collaborative framework in which external developers may develop plug-ins
collaborating with each other, such a well designed architecture even becomes
more important.

In this experience report, we explain difficulties of developing and
maintaining open extensible collaborative analysis frameworks, through the
example of \framac, a platform dedicated to the analysis of code written in
C. We also present the new upcoming software architecture of \framac and how it
aims to solve some of these issues.
\end{abstract}


\section{Introduction}

This experience report is about \textbf{software architecture}. Software
architecture may be defined as the set of structures needed to reason about the
system which comprise software elements, relations among them, and properties of
both~\cite{bachmann10}. It has long been recognized that implementing a large
piece of software without a well-engineered software architecture will stumble
along or, most likely, fail~\cite{bass03}. Thus, whatever the software is, its
architecture does matter. This article aims at providing a practical evidence
about this statement through a concrete example.

This short paper is not only about software architecture. It is also about
\textbf{code analysis tools and formal methods} based tools. Code analysis is
known to be complex and difficult. In particular when based on formal
methods. Besides formal methods are still a very active research domain in which
new techniques and improvements are discovered frequently. Consequently tools
based on such techniques evolve quickly, or become outdated quickly. Despite
this harsh reality, some analysis tools become mature enough to be used in
industrial settings. Such tools must combine the intrinsic hardness of code
analysis with the intricacies of industrial code they aim to analyze. Thus they
are large and complex pieces of software. So they need well-engineered software
architectures which must permit tools to fulfill two opposite requirements: be
\textbf{flexible} enough to allow rapid prototyping and experimentation; be
\textbf{stable} enough to obtain reproducible predictable results and no
regression after upgrading.

Few code analysis tools are code analysis \textbf{frameworks}. Frameworks
can be characterized by few key features. First \textbf{extensibility} which
allows to extend the framework with new features. Then \textbf{collaboration}
which allows to mix extensions to provide new super features quickly and
easily. Next \textbf{inversion of control} which gives the control of the
overall application to the framework itself and not to the application that uses
it~\cite{johnson88}. Finally \textbf{opening}, usually \emph{via} open-sourcing,
which allows external developers to contribute.

Perhaps not surprisingly, developing and maintaining a large flexible yet mature
open extensible collaborative code analysis framework is not that easy. But
thankfully a software architecture may be very helpful if carefully
designed. The goal of this short paper is to provide evidence about these
statements, through the example of \framac~\cite{fac15}, which is such a
framework dedicated to the analysis of code written in \C.

Section~\ref{sec:fc} quickly introduces \framac. Section~\ref{sec:req} details
requirements which must be fulfilled to provide the framework key
features. Section~\ref{sec:archi} presents the evolutions of the \framac
architectures up to the upcoming one and explains which requirements they try to
address.

\section{\framac at a Glance}\label{sec:fc}

\framac~\cite{fac15} is a platform dedicated to the analysis of source code
written in \C. The \framac platform gathers several analysis techniques into a
single collaborative extensible framework. For instance it contains \Value,
which is a value analysis by abstract interpretation~\cite{value}, \WP, which is
a program proof tool through weakest precondition calculus~\cite{wp}, and \eacsl
which is a monitoring tool~\cite{sac13}.

The analyzers share a common specification language called \acsl~\cite{acsl}:
any analyzer can prove the (un)validity of each \acsl annotation of a \C
program, but they can also generate them. That is one way to ensure
\textbf{analyzer collaboration}. For instance the \rte tool~\cite{rte} is able
to generate an annotation for each potential runtime error, and then \Value, \WP
or \eacsl can try to prove each of them, while \framac itself combines what is
proven by any single tool to compute what still remains to be proven on the
whole~\cite{fmics12}. Another way to allow analyzer collaboration is through
the use of their \ocaml programmatic interfaces (\emph{aka} APIs) by each
other. For instance, the callgraph may use the results of \Value to resolve
function pointers.

\framac is also \textbf{extensible}: it is possible to develop new analyzers in
\ocaml and to integrate them in the framework. Anyone can do this thanks to its
licensing policy (LGPL, version 2): \framac is \textbf{open source} and most of
its plug-ins as well (for instance all of the above-mentioned plug-ins are open
source), while it is still possible to implement close-source proprietary
extensions~\cite{taster,fanc}. This way, \framac is not only a collection of
collaborative tools but also a \textbf{development platform} in \ocaml which
targets both \textbf{academic and industrial} users and
developers~\cite{cuoq:icfp09}.

\section{\framac Requirements}\label{sec:req}

The \framac overview induces nice properties that the tool must have. First,
\framac must be usable both from the command line and in a graphical user
interface (GUI). Second it must analyze (ISO 99) \C code annotated with \acsl
specifications. Consequently it must provide a \textbf{\C front-end}, extended
by an \textbf{\acsl implementation}. Particularly it must provide at least one
abstract syntax tree (AST) from annotated \C files. Related to these
requirements, \framac must be easily identified as a single tool from a user
point of view. That means primarily \textbf{homogeneity}. For instance, user
interactions like getting inputs or printing messages should be as uniform as
possible in order to make the learning of the tool easier and faster, while it
must be possible to integrate new analy\-zers in the GUI smoothly. However
\framac is also a collection of possibly very different tools: using a
monitoring tool based on program transformation techniques is not the same as
using a program proof tool and both are different from an abstract interpreter
in many aspects. This \textbf{heterogeneity} of uses means that \framac must be
\textbf{customizable} enough to fit the analyzer needs. Being both
\textbf{homogeneous} and \textbf{heterogeneous} while being user friendly
implies that the user should identify quickly what is common to all analyzers
and what is specific to each of them. For instance, a code analysis tool has
usually many parameters which allow the user to tweak the tool according to its
current use case. In our context, it must be able to quickly know which
parameters are common to all analyzers (\emph{e.g.} the hypotheses used to
compute system-dependent information like the size of \C types) and which ones
are relevant only for a specific (set of) analyzer(s) (\emph{e.g.} the slicing
criteria used by the slicing tool).

Analyzer collaboration is provided by two means. The first one is exchanging
\acsl annotations: analyzers must be able to generate new annotations when
required (for instance to make explicit their correctness hypotheses) which
several other analyzers could try to verify. In this context, the kernel ensures
global \textbf{consistency} by consolidating which property is proved by which
analyzer under which hypotheses~\cite{fmics12}. The second way of analyzer
collaboration is through APIs: a plug-in may directly use values (including
functions) exported by others plug-ins either to tweak their behaviors or to
compose them to quickly develop powerful analyzers.

In order to ensure homogeneity and inversion of control, \framac itself must
control the overall execution of the tool and not let each individual analyzer
decide. For instance, \framac must parse itself the command line, display
messages, run the GUI, parse the source code, run the analyzers,
\emph{etc}. However it must let place to analyzer customization when it makes
sense.

Furthermore \framac must be not only a tool but also a large library which must
provide useful services to \textbf{ease analyzer developments}. It must be
developed in \textbf{\ocaml}~\cite{cuoq:icfp09}.


\section{\framac Architectures}\label{sec:archi}

While the first \framac public release was in 2008, \framac is developed
since~2005 and 3 different architectures have been successively designed. The 
prehistoric one, presented in Section~\ref{sec:prehistoric}, was used from the
beginning of the development until 2007. It was then replaced by the current
architecture, detailed in Section~\ref{sec:current}. It is now in turn going to
be replaced by the upcoming one explained in
Section~\ref{sec:upcoming}. Section~\ref{sec:cost} estimates the cost of these
evolutions. 

\subsection{Prehistoric Architecture}\label{sec:prehistoric}

\begin{figure}[htbp]
\begin{center}
\includegraphics[scale=0.8]{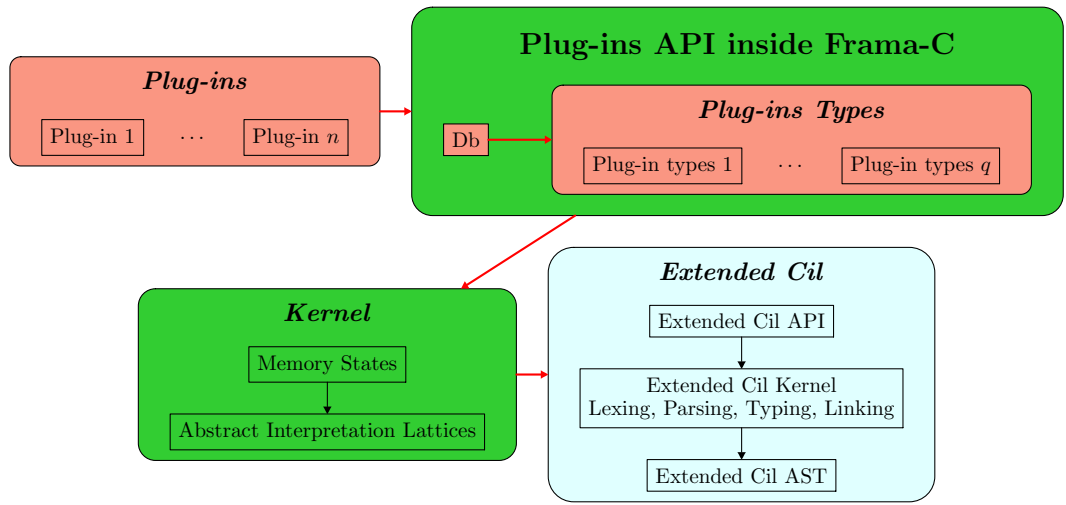}
\caption{\framac Prehistoric Architectural Description}\label{fig:prearchi}
\end{center}
\end{figure}

The \framac architecture is from the origin based on a single fundamental
principle: it is a plug-in oriented architecture \emph{\`a la}
\eclipse\footnote{\url{http://eclipse.org/}} in which each analyzer is a plug-in
which uses a kernel. This design choice is nowadays the standard way to provide
extensibility while helping to solve the homogeneity \emph{vs} heterogeneity
issue: the kernel ensures some homogeneity while plug-ins may develop their own
concepts when required. Analyzer collaboration through API is ensured by the
kernel which maintains a plug-in database storing the API functions. However, at
the very beginning of the \framac development, this principle was more an
essential principle than a perceptible reality expressed by
Figure~\ref{fig:prearchi}.

Indeed plug-in developers have to modify themselves some parts of the kernel
(particularly the plug-in database and the Makefile) to register new analyzers,
while it was not possible to plug or unplug them without recompiling the whole
tool. More generally, there were no clear distinction between plug-ins and the
kernel. There were at least two major reasons for such issues. First \framac was
initially developed in a hurry by a couple of persons to demonstrate that it
could be a viable project which it is worth being founded. Second \ocaml had no
native dynamic linking at that time and thus a ``plug-in'' has to be statically
linked against the rest of the platform. Also \ocaml does not support mutually
dependent compilation units: that complicates a bit the overall organization.

This plug-in oriented architecture naturally leads to inversion of control: the
kernel controls the \framac execution and decides by its own when each part of
each plug-in must be executed. For instance, one part of a plug-in is ran when
parsing the command line to handle plug-in specific options and another part is
ran to execute the main function of the analyzer.

The other main design choice of the prehistoric architecture is the use of
\cil~\cite{cil} as \C front-end. \cil is also both a tool and a library but
\framac only uses the library part. It provides a \C AST from \C source files
and an API to use it easily. This API is almost organized in a centralized way:
a single large compilation unit (\texttt{Cil}) contains almost the whole API. At
that time, \framac directly reuses \cil as such, but extended it in order to
support \acsl. Even if \framac integrated its own version of \cil, it was
regularly synchronized with the Berkeley mainstream version. Finally, since the
very first important plug-in was \Value, the main library provided by the kernel
was an abstract interpretation toolkit.

\subsection{Current Architecture}\label{sec:current}

\begin{figure}[htbp]
\begin{center}
\includegraphics[scale=0.8]{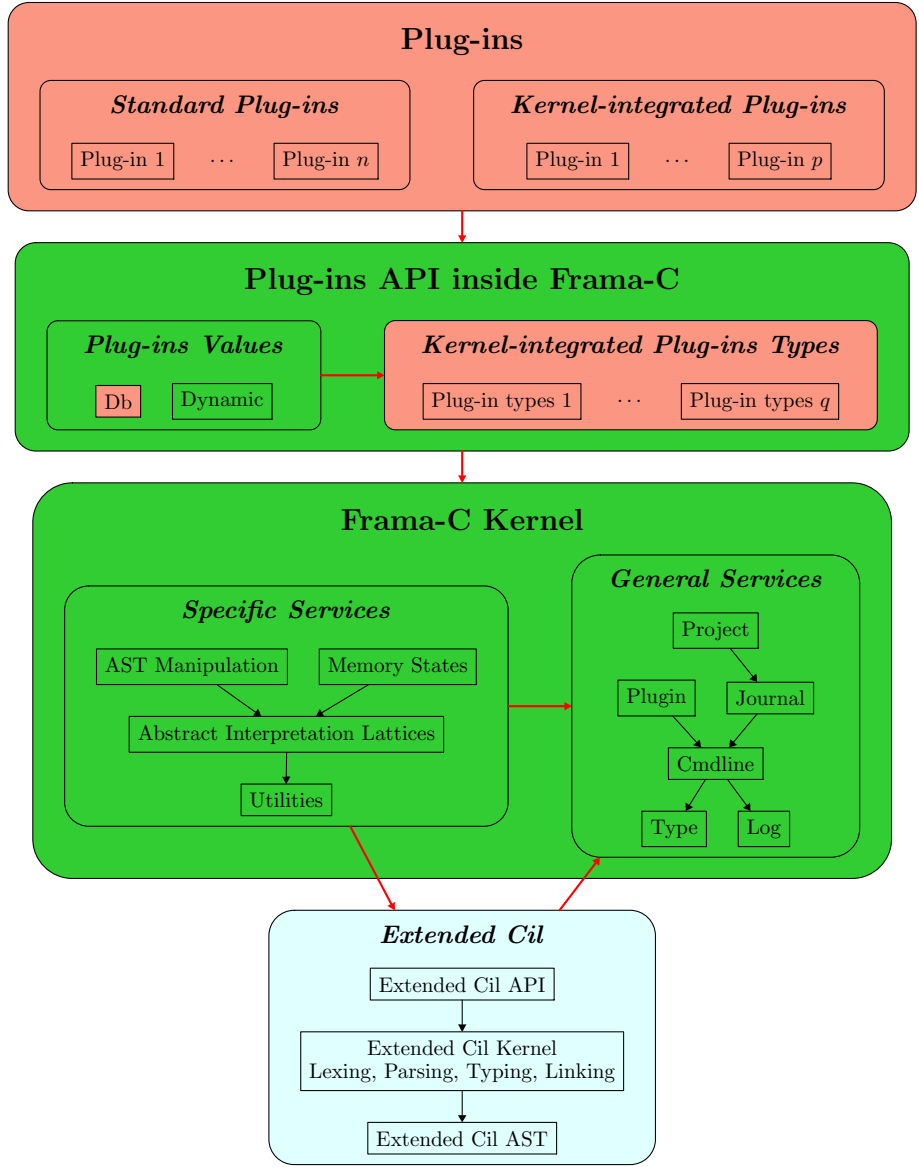}
\caption{Current Architectural Description}\label{fig:old-architecture}
\end{center}
\end{figure}

Figure~\ref{fig:old-architecture} introduces the \framac architectural
description which was defined in 2007. It is still in use in the last open
source release of March 2015 (namely \emph{Sodium}). When defined, it was
explained in a document which more generally explains how to develop \framac
plug-ins~\cite{plugin-dev-guide}.

Thanks to the introduction of native dynamic loading into \ocaml, it becomes
possible to provide fully independent plug-ins developed outside the platform
without any modification of the kernel nor recompilation, even if the old ways
of developing plug-ins is still supported. However \ocaml is a statically typed
programming language. That leads to technical issues with APIs, packing
facilities\footnote{An \ocaml pack is a way to group together at compile time a
  set of compilation units into a single one. In \framac, each plug-in is
  compiled this way in order to not pollute the whole namespace with the names
  of the plug-in internal compilation units.} and dynamic loading which limit
plug-in collaborations through API to simple uses. In particular, registering
new types is limited to abstract types, while registering values is restricted
to monomorphic values and remains tedious through the use of a dedicated
library~\cite{signoles:jfla11}. In practice, this limitation is tractable as
long as no plug-in needs to export a large set of functions~\cite{cuoq:icfp09}.

Within this architecture, the kernel is still based on \cil but it also provides
its own services as libraries in order to simplify plug-in development and to
ensure plug-in consistency. Just after the introduction of this architecture,
\cil itself had been fully forked from Berkeley's mainstream: in addition to the
support of \acsl, the changes introduced to be consistent with the other parts
of the kernel (unified ways of displaying messages and handling errors,
integration of the project system~\cite{project}, \emph{etc}) were so huge that
synchronization with the Berkeley's version became intractable.

Inversion of control is an important feature of a framework, but a plug-in often
needs to customize the default behavior. Such an \emph{a posteriori}
modification is not so easy to provide in a statically typed functional language
like \ocaml. It is usually done by hooking which consists in registering a
closure to be executed at a well defined moment of the execution. Yet these
hooks do not appear as such in the architectural description. They have
nevertheless a major impact in the overall organization of the framework since
they get back some control to plug-ins at many places.

This architecture fulfills most requirements from plug-in collaborations to
extensibility through heterogeneity of uses \emph{via} customizability and
global homogeneity of the platform \emph{via} inversion of control. It probably
contributes to the adoption of \framac throughout the world. \framac and its
community -- including plug-in developers -- are still growing: the number of
line of codes (\emph{loc}) of the open source version as well as the number of
plug-ins and developers (including third-party ones) constantly increase.

However that has revealed an important issue of the platform: non expert plug-in
developers are often not able to find what they are looking for in the \framac
API, and even do not know where to search in. Even expert developers working on
\framac since a long time face such difficulties regularly. This annoying
situation is partly inherent to 100+-kloc libraries: searching one piece of
information in a large code base is never easy, particularly when
unknown. However, in our case, the quality of the kernel APIs also regularly
deteriorate because it becomes more and more unclear where to add a new function
or a new library. For instance, basic functions to manipulate the AST are
dispatched in several files of different directories without any apparent logic.

Indeed huge libraries are usually organized in a decentralized way: they are
split into lots of files, small enough to remain tractable. \framac has been no
exception to the rule while growing. However this organization clashes with the
\cil centralized organization. Thus a large part of the kernel has no clear
organization at all and no one is able to explain where is the best place to
add. For instance, should a new function about some ACSL construct be added in
\cil among plenty of other functions, or in a \cil extension, or in some other
kernel files dedicated to ACSL? When it is unclear where to add a function, it
is \emph{a fortiori} unclear where to search for a value. Furthermore some
(parts of) modules only address very specific needs or even should only be used
by the kernel and does not target the standard plug-in developer. In particular,
there are plenty of forward references in the kernel just to circumvent the
absence of mutually recursive compilation units in \ocaml. However they remain
visible by anyone. These drawbacks contribute to the current issue.

Another issue is that \framac is still in heavy development because it tries to
remain at (or to reach) the cutting edge of scientific research: that leads to
unavoidable instability of several parts of the framework. Of course it is
annoying for any developer which has to maintain its plug-ins, even if we
provide as far as possible automatic migration scripts.

Finally the last issue is that few plug-in developers begin wanting to provide a
large API with tens of functions and types for their plug-ins and it is almost
intractable with the current limitation of dynamic loading. For instance, one
use case is to develop a plug-in whose the main goal is to be a dedicated
library for other plug-ins.

\subsection{Upcoming Architecture}\label{sec:upcoming}

\begin{figure}[htbp]
\begin{center}
\includegraphics[scale=0.8]{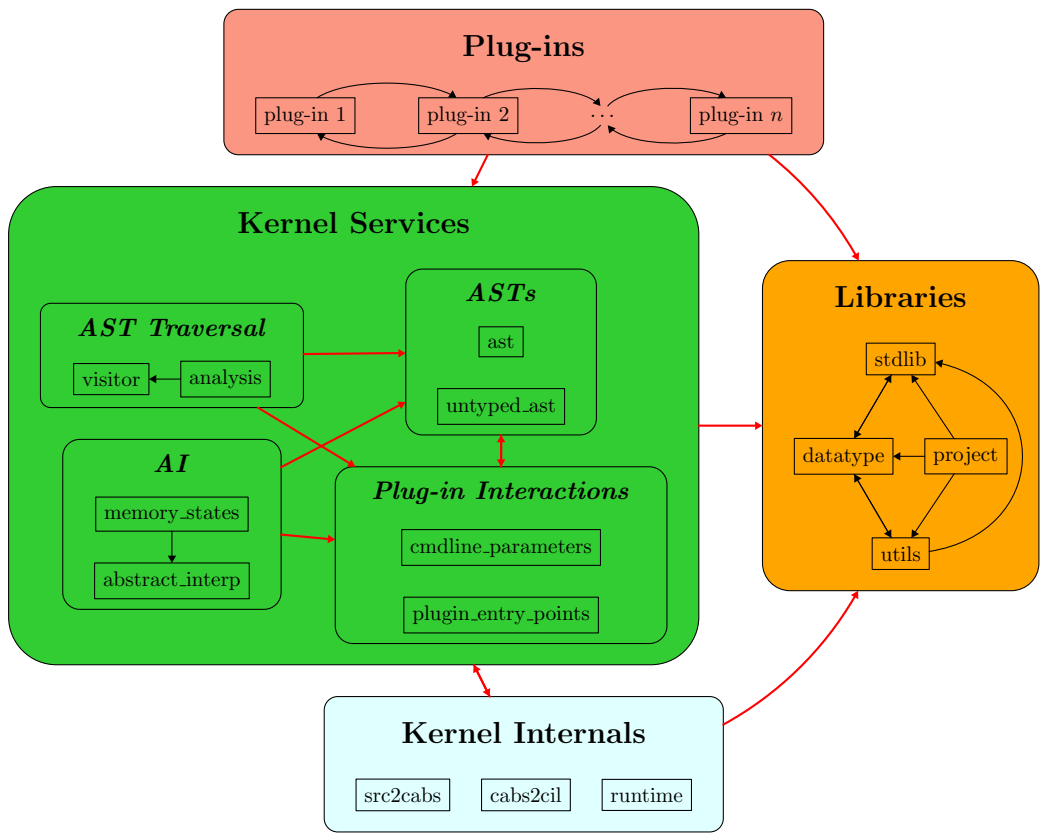}
\caption{Upcoming Architectural Description}\label{fig:architecture}
\end{center}
\end{figure}

To synthesize the previous section, the current architecture almost fulfills our
needs, but we hugely need to help the developer to find what is looking for by
providing a way to quickly and precisely know what every module and single
function is useful for. Namespace is a standard way to address this issue at the
programming language level. However \ocaml has a poor support of namespaces even
if there are currently intensive discussions around this topic in the \ocaml
community: for instance it is not possible to indicate that a value or a module
can only be used by a single family of values or modules (like \cpp's friendship
or \java package visibility for instance). Therefore we decide to address this
issue at the architecture level. This choice leads to the architecture presented
Figure~\ref{fig:architecture} which is currently deployed in a \git branch of
the private development version of \framac. It is planned to be part of the next
public release.

The main design idea of this new architecture is to split \framac into distinct
areas, clearly separated thanks to the directory structures. First plug-in
(sub-)directories delivered with \framac are no longer in the same directory
than most kernel directories, but are group together in another directory. So
there is a clear boundary between the kernel and the plug-ins which is visible
at a glance by any developer. Second the kernel itself is split into three main
\hyphenation{areas}areas: Libraries, Kernel Services and Kernel
Internals. \emph{Libraries}
contains modules which are not specific to code analysis and could be moved
outside \framac easily (\emph{e.g.}  extension of the \ocaml standard libraries,
general datastructures like bags or bit vectors, wrapping of system commands,
\emph{etc}). \emph{Kernel Services} provides useful services to plug-in
developers (\emph{e.g.}  access to the AST, abstract interpretation toolkit,
standard interaction points between plug-ins and the kernel,
\emph{etc}). \emph{Kernel Internals} contains modules which should be useless
for plug-in developers (except for very advanced uses reserved to experts
through hooking) like the generation of the ASTs from a \C source code or the
routines run when initializing \framac. This design idea should almost solve the
previously identified issues by reducing and easily identifying the place to
look for when searching something in the API. For instance, no need to search in
the kernel internals anymore, while a developer may not even easily know that a
module should not be used outside the kernel in the previous architecture.

A major consequence of this new architecture is to remove \cil: even if the
implementation is still the same, it does not appear anymore in the whole
architectural description. Yet deploying the new architecture is a two-step
stage and only the first one is being finalized. It only modifies the directory
level, without changing the module APIs. Thus it has no impact on the existing
plug-ins based on the \framac API. The second step is still not precisely
planned in our timeline but it will remove the \cil centralized organization by
modifying \framac API. In order to ease the migration from old versions to the
newest one, we will design new services one at a time from existing
functionalities which are currently dispatched at several places of \cil. Then
\cil will be removed when it will be rendered meaningless and small enough. Past
experiences with the (re)design of other \framac services have already shown
that it is often the less painful way to introduce major incompatibilities
between releases. It will eventually solve the \cil-centralized \emph{vs}
kernel-decentralized issue. It could be pointed out that it could deserve the
adoption of the platform by existing \cil users, but mainstream \cil and the
forked \cil version of \framac have diverged enough to already be an issue with
the current architecture. Also the benefits for all the \framac developers
community should be much larger than this potential drawback for a small part of
the community.

Finally the last action to help developers to find out about what they are
looking for is to improve what we could call the micro software architecture:
API organization in a single compilation unit. As already explained, such a unit
often contains hooks, parts dedicated to kernel or advanced uses, (almost)
stable parts or parts which could still evolve in the near future,
\emph{etc}. At this level, we are continuously updating the API to clearly
separate and document each part in order to indicate the functionalities from
the more idiomatic ones to the niches one. For instance, we put at the very end
of the module API the kernel-only features and they are even hidden in the
generated \textsl{HTML} documentation.

The other evolution which comes from this new architecture is the interactions
between plug-ins. The limitations of plug-in APIs almost disappeared thanks to
\ocaml's first class modules\footnote{They were first included in \ocaml 3.12.0
  released on January 2012, but major evolutions in recent versions of \ocaml
  makes them much easier to use.}. This \ocaml evolution allows \framac to
replace the previous heavy interaction mechanism through a standard
communication \emph{via} file interfaces. There is still room for improvements
here, but the previous mechanisms are nowadays almost deprecated, and their uses
will be gradually replaced. It will eventually allow plug-ins to provide an API
as large as they wish and thus to solve the last issue mentioned in
Section~\ref{sec:current}.

\subsection{Estimation Cost of the Changes}\label{sec:cost}

Estimating the manpower required to deploy the current and the upcoming
architecture is not that easy mainly because it depends on what is counted and
what is not counted. Here I do not take into account the development of new
services like dynamic loading or the project system, but I do take into account
file modifications (e.g. changes in Makefiles) required by the architectural
changes.

The current architecture was implementing in 2007 by almost a single developer
(the author of this article) in few weeks over several months. Most efforts was
spent in clearly separating the plug-ins from the kernel in Makefiles and
configures' scripts. Modifying the directories and files structure was not a big
deal because \framac was still in its early days at that time: there were not
that many plug-ins nor developers.

Implementing the upcoming architecture is a more important effort which is still
not ended. We have already spent several men-Weeks of developments and intensive
discussions over 5 months just to define the architecture directory-level
without talking so much about the file-level and the process is only closed to
be finished in its first version. This large amount of time comes from there are
more core developers than before and they may have divergent opinions. More
importantly, the development is done in a separated \git branch along several
months since no developer have enough time to only work on this topic. This
branch only consists in creating and deleting directories and moving files. As
expected, it has no impact on API and so on plug-in developers. However it may
conflict a lot with other branches which do the same kind of operations since
\git is really bad to merge two branches if both moves the same file in
different ways (or if one branch moves the file and the other one deletes
it). Unfortunately these patterns already occur several times (for instance, one
merged branch reorganized non-regression test directories of a major
plug-in). There were also heavy changes in the Makefile which conflicted with
the Makefile changes of the branch of the new architecture. At the end, heavy
\git conflicts happens much more frequently than expected.

The second stage of the new architecture deployment, which will modify the file
APIs, is still not precisely planned in our timeline but it will be done part by
part along months or even years of sparse developments on that topic.

\section{Conclusion}

Through the evolution of the \framac architecture, this article has explained
why software architecture matters. In particular, it has shown how software
architecture may help to fulfill requirements of a code analysis frameworks and
thus have a major impact in the daily life of its developers, including third
party ones. 
Software architecture is only an element over several other engineering and
technical choices. For instance, documentation is also crucial. Also software
architecture is not only a black board exercise. It has a concrete
representation which must take into account technical issues like the choice of
the underlying programming language. Finally, the software architecture must
evolve along with the whole software lifecycle and potentially be replaced when
becoming outdated. However such a change has an huge impact on the whole
software and must be made with care in order to limit annoyance. In this
respect, change management techniques are useful to help the developers to
migrate their code as easily as possible without mumbling too much, to
understand the new architecture and eventually to accept such a major change.

An open plug-in based architecture is the key of the adoption of a tool like
\framac by both the academic and the industrial formal methods
communities. However it is not so easy to deploy in practice and the sooner the
better. Also, while reusing existing libraries saves a lot of time, they must be
redesigned to be well integrated into the architecture as soon as it has to be
extensively modified (like \cil in \framac).

\section*{Acknowledgments}

Special thanks to Virgile Prevosto and Boris Yakobowski who spend a large amount
of their time to review the huge \textsl{Git} branch implementing the new
architecture. I would also like to thank Fran\c{c}ois Bobot and Lo\"ic Correnson
for their constructive feedback and suggestions about it as well as Florent
Kirchner who provides a continued support to this major change. Furthermore the
anonymous referees contribute to improve the final version of this paper thanks
to their fruitful remarks.

\bibliographystyle{eptcs}
\bibliography{./main}

\end{document}